\documentclass[preprint2]{aastex}
\usepackage{graphicx}

\slugcomment{Accepted in ApJ}
\lefthead{S. Takagi et~al.}
\righthead{X-ray Sources in the Sgr~B2 Cloud Observed with Chandra}

\begin{document}

\title{X-ray Sources and Star Formation Activity in the Sgr~B2 Cloud
Observed with {\it Chandra}}
\author{Shin-ichiro~Takagi, Hiroshi~Murakami and  Katsuji~Koyama}
\affil{Department of Physics, Faculty of Science, Kyoto University,
Sakyo-ku, Kyoto 606-8502, Japan; takagi9@cr.scphys.kyoto-u.ac.jp,
hiro@cr.scphys.kyoto-u.ac.jp,~koyama@cr.scphys.kyoto-u.ac.jp}

\begin{abstract} 

 We report the X-ray population study in the giant  molecular cloud
Sagittarius B2~(Sgr~B2). 
More than  a dozen of X-ray  cloud members (and candidates) are
discovered with {\it Chandra}.
Two bright  X-ray sources are located near Sgr~B2 Main, the most copious
complex of the ultra compact HII sources. 
The X-ray spectra are fitted with a thin thermal plasma model
of 5--10 keV temperature.  
The intrinsic luminosity after correcting the
absorption of $\sim5\times10^{23} \rm ~H~cm^{-2}$ is 
$\sim10^{33}$ erg s$^{-1}$.
Although these two X-ray sources are attributable to 
young stellar objects (YSOs) in the same HII complex, they are in sharp contrast; 
one at the  center of the HII complex  exhibits strong K-shell
transition lines of iron, while the other near the
east  has only weak lines.  
The other HII complexes, Sgr~B2 North and South,
also show hard and highly absorbed X-ray emissions due possibly to 
the star formation activity.

The composite X-ray spectrum of the other cloud member X-ray sources
is fitted with a thin thermal plasma of $\sim$10-keV
temperature with the hydrogen column density ($N_{\rm H}$) of $1.3\times10^{23}
\rm ~H~cm^{-2}$, and the individual X-ray luminosity of a few times
of 10$^{31-32}$ erg s$^{-1}$. These are likely to be a single or cluster of
YSO(s), but neither radio nor infrared counterpart is found.
An alternative scenario of isolated white 
dwarfs powered by the Bondi-Hoyle accretion from the dense cloud gas is also discussed.
The X-ray spectra  exhibit  an additional 6.4-keV line of neutral
or low-ionization irons, which indicates that  the environment
gas is concentrated near at the sources.

\end{abstract}

\keywords{Galaxy: center --- ISM: clouds ---
ISM: individual (Sagittarius B2)---
stars: formation ---
X-rays: individual (Sagittarius B2)---
X-rays: stars --- X-rays: spectra
}

\section{Introduction} 

The giant molecular cloud Sgr~B2, located at the projected distance of 
$\sim$100 pc from the Galactic center (GC), is one of the richest star forming 
regions (SFRs) in our Galaxy.
Due to its proximity to the GC, Sgr~B2 is heavily obscured in the optical,
even near infrared (NIR) and soft X-ray bands.  
Accordingly, the star formation activity has been traced
mainly with  the radio and mid--far infrared (MIR and FIR) bands.  
The  VLA continuum observations have revealed at least 12 separate HII 
regions in the Sgr~B2 cloud (Benson \& Johnston 1984), which are
grouped into three main concentrations, Sgr~B2~North (HII region K), 
Main (HII regions A--G, I, and J) and South (HII region H).
The follow-up high-resolution continuum observations have revealed nearly
60 ultra  compact (UC) HII regions lying along  the north-to-south
elongation (Gaume et~al.\ 1995; De Pree, Goss, \& Gaume 1998), which are
also traced by molecular  lines and FIR (e.g. Sato et~al.\ 2000; Goldsmith
et~al.\ 1992).
Clusters of molecular maser sources are found from the HII complexes,
Sgr~B2~North~(N), Main~(M), and South~(S) 
($\rm H_{2}O$: Kobayashi et~al.\ 1989; OH:
Gaume \& Glaussen 1990; SiO: Mehringer, Goss, \& Palmer 1994;
Shiki, Ohishi, \& Deguchi 1997).
The ages of Sgr~B2~(N) and Sgr~B2~(M) are estimated to be  
$\sim10^4$ years, using the photon flux of the exciting stars and
the ambient gas density (De Pree et~al.\ 1995, 1996).  
Powerful bipolar outflows are found from Sgr~B2~(N)
and Sgr~B2~(M), with the dynamical age of  $\sim 10^4$ years (Vogel, Genzel,
\& Palmer 1987; Lis, et~al.\ 1993; Mehringer 1995; Kuan \& Snyder 1996; 
Liu et~al.\ 1998). 
MIR sources are also found at Sgr~B2~(M) and (S) (Egan~et~al.~1999).
All these results indicate that Sgr~B2 harbors  many clusters of  
very young stellar objects (YSOs) of  high-mass stars.

There is mounting  evidence that  star forming regions, whatever low-mass
or high-mass, generally emit fairly strong  X-rays 
(e.g., Feigelson \& Montmerle 1999 and references there in;
Garmire et~al.\ 2000).  
$ASCA$ and later $Chandra$ revealed that a significant fraction
of  YSOs emit hard X-rays above 2~keV 
(e.g., Koyama et~al.\ 1996a; Shultz et~al.\ 2001; 
Kohno, Koyama. \& Hamaguchi 2002). 
The hard X-rays can penetrate through the large Galactic $N_{\rm H}$ to
Sgr~B2, hence may provide a new view on the star formation  activity 
in this cloud.
This paper reports on the first $Chandra$ results of hard X-ray sources and
studies the X-ray population in Sgr~B2.

Throughout this paper, the distance to Sgr~B2 is assumed to be
8.5 kpc, the same as to Sgr~A$^*$, and is similar to the estimated
distance of 7.1 $\pm$ 1.5~kpc \citep{Reid88}.

\section{Observations and Analyses} 

\subsection{Source Detection and Identification} 

The $Chandra$ ACIS observation of Sgr~B2  was carried out  on 29--30 March
2000 with the nominal aim-point at $\rm R.A.~(2000) = \rm
17^{h}47^{m}07^{s}$, $\rm Dec.~(2000) = \rm -28^{d}26^{m}29^{s}$. 
The satellite and instrument are described by Weisskopf, O'dell, 
\& van Speybroeck
(1996)  and Garmire, Nousek, \& Bautz (2001), respectively.
Sgr~B2 lies near the center of the ACIS-I array of  four front-side
illuminated CCDs, each with  $1024 \times 1024$  array of  $0\arcsec.5
\times 0\arcsec.5$ pixels covering a $8\arcmin \times 8\arcmin$ field of
view.  
Using the Level~2 processed events provided by the pipeline processing 
at the $Chandra$ X-ray Center, we selected  the $ASCA$
grades\footnote{see http://asc.harvard.edu/udocs/docs/POG/MPOG/index.html}
0, 2, 3, 4, and 6, 
as X-ray events; the other events, which are due to charged particles,
hot and flickering pixels, are removed. 
The effective exposure is about 100 ksec.

In the soft (0.5--2.0~keV) band image of the ACIS-I field of $17\arcmin.4
\times 17\arcmin.4$,
we find four bright X-ray sources that have {\it Tycho-2} counterparts.
The positions of the {\it Tycho-2} sources have accuracy of 60~mas 
\citep{Hog00}, hence the {\it Chandra} frame is
fine-tuned using these four sources within an absolute error of 0.25 arcsec
(1~sigma dispersion).
In order to search the Sgr~B2 cloud member sources, we run the program of
{\it wavdetect}
\footnote{see http://asc.harvard.edu/udocs/docs/swdocs/detect/html/} in the
2--10~keV band, because X-ray photons below  2~keV  are totally absorbed
by the interstellar gas to Sgr~B2 of $\sim 1 \times 10^{23} \rm
~H~cm^{-2}$ (Sakano 2000).
The threshold significances of the {\it wavdetect} are set at $10^{-6}$
and 0.001 for the source list and the
background estimation, respectively. 
The wavelet scales are 
1, $\sqrt{2}$, 2, $2\sqrt{2}$, 4, $4\sqrt{2}$, 8, $8\sqrt{2}$, and 16
pixels.
We then find 15 sources in the Sgr~B2 cloud area of a $3'\times3\arcmin.5$ region.

\begin{figure}
 \figurenum{1}
 \epsscale{0.8}
  \plotone{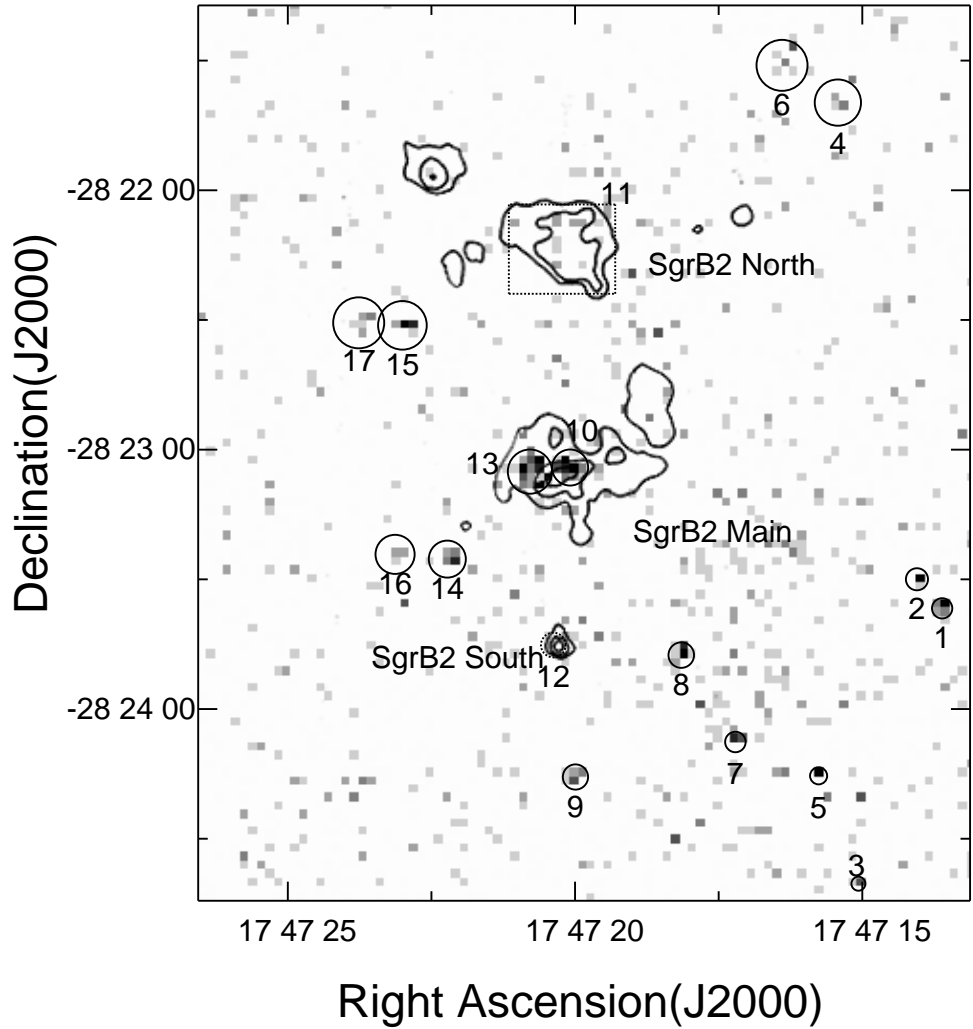}
 \figcaption[f1.eps]
{The 2--10keV  X-ray band image of  Sgr~B2, overlaid on contours of HII regions 
~(Gaume et~al.\ 1995). 
The bin size is $ 1\arcsec \times 1\arcsec $~($ 2\times 2$ pixels are summed). 
The solid circles and dotted area~(box: No.~11, circle: No.~12) indicate the
source regions  for the  X-ray sources detected with '{\it wavdetect}' and manually
, respectively~(see text). \label{fig;b2all}}
\end{figure}

The 2--10~keV band image of this region and the position of detected X-ray
sources are given in Figure~\ref{fig;b2all} by the solid  circles with 
a diameter of three times of the half power diameter (HPD) of the 
point spread function (hereafter, $3\times$HPD).
In addition to the {\it wavdetect} search,  we
inspect X-ray sources manually from all the compact HII regions including
the Sgr~B2~(N) and (S) complexes. For this search, we extract the X-ray
photon from a $3\times$HPD-diameter circle or from the same area
as the HII regions (those  extending  larger than $3\times$HPD).
We then find  X-ray excess from  Sgr~B2~(N) and Sgr~B2~(S) with more than
$3\sigma$ confidence  level (in the 2--10~keV band).
The source areas  are also given by the dotted box (Sgr~B2~(N)) 
and  circle (Sgr~B2~(S)) in Figure~\ref{fig;b2all}. 
Since this paper focuses on single or compact cluster of point sources,  we 
do not search diffuse sources  other than those of the HII regions.
All the detected sources,  the positions and photon counts are listed in Table 2.

\begin{figure}
 \figurenum{2}
 \epsscale{0.8}
 \plotone{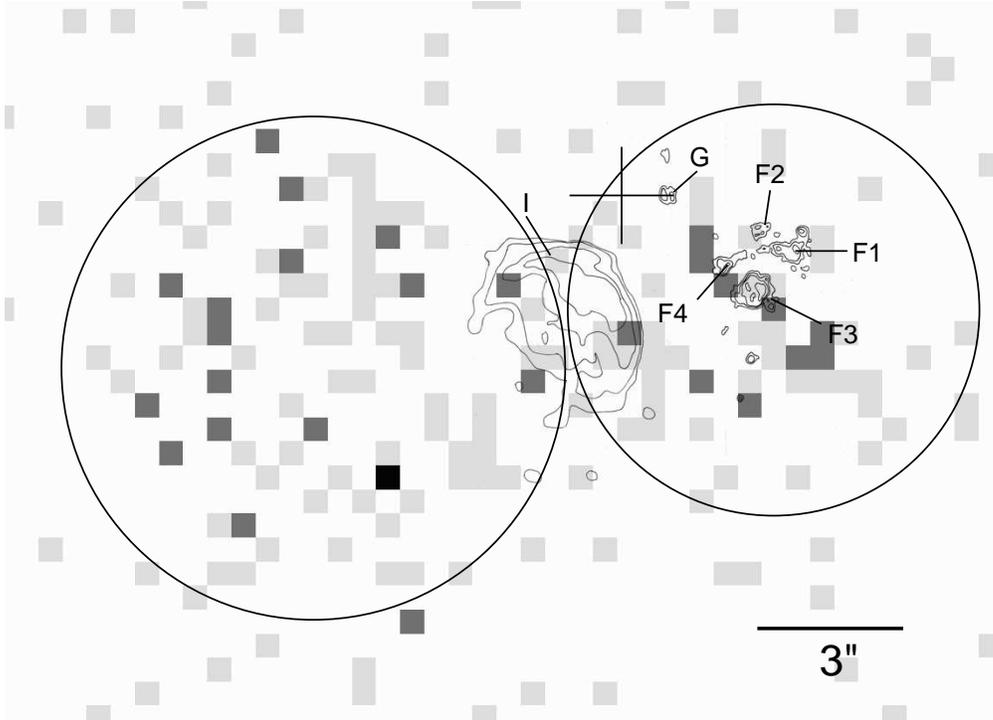}
 \figcaption[f2.eps]
{The 2--10 keV band X-ray image. near Sgr~B2~(M), overlaid on 
the contours of HII regions F1--F4, G~(De Pree et~al.\ 1998),
and I~(Gaume \& Claussen 1990). 
The bin size is  $0.\arcsec 5 \times 0.\arcsec 5$.
The cross~(+) indicates the position and $1\sigma$ errors~($\sim 1.\arcsec 1$)
of the MIR source MSX5C~G000.6676-00.0355.
The right source is No.~10, and the left source is No.~13.
The X-ray spectra~(see Figure~\ref{fig;mainspec}) are
obtained from the solid circle regions~(see text).
\label{fig;mainimage}}
\end{figure}

The closed-up image of Sgr~B2~(M) is shown in
 Figure~\ref{fig;mainimage} overlaid on  the radio contours. 
The brightest X-ray sources in Sgr~B2, No.~10 and 13 are located in this complex.

\subsection{Spectrum and Time Variability} 

The X-ray spectra are made from the source areas given in Figures 1 and 2.
In order to properly subtract the diffuse X-rays in the Sgr~B2 and GC
regions (Murakami, Koyama, \& Maeda 2001; Koyama et~al.\ 1996b), 
we make the background spectrum  from the full region of Sgr~B2  
excluding all the source areas (see Figure~\ref{fig;b2all}).
Since No.~10 and 13 are located in the local diffuse maximum near 
the HII complex  Sgr~B2~(M) (Murakami et al.~2001), the background data
are taken  from  a  $30\arcsec$-radius circle with the center at the
middle of the two sources (the data in the source regions are excluded).
No.~16, on the other hand,  lies in a  local diffuse  minimum, 
hence the background is taken from an annulus around the source.

To eliminate obvious foreground or background sources,  
the background-subtracted spectra are fitted with  a model either
power-law of fixed photon indices 2.0, 
or 1-keV thin thermal plasma of solar-abundances; the former is
representative to background AGNs and the latter is for foreground stars.  
The interstellar $N_{\rm H}$ to Sgr B2 is
$\sim 10^{23}$~H~cm$^{-2}$ (Sakano 2000), while that of  
intra-Sgr B2 cloud
is $\sim10^{24}$~H~cm$^{-2}$ (Lis \& Goldsmith 1989). 
Thus the sources well outside  the
(1--10)$\times10^{23}$~H~cm$^{-2}$
range  should be regarded  as either background AGNs or foreground
stars.
No.~2 and 5 show the $N_{\rm H}$ of $<$ 1
$\times10^{23}$~H~cm$^{-2}$ within the error, hence are regarded
as foreground stars. The others
are likely located in the cloud (hereafter, the cloud members).

\begin{figure}
 \figurenum{3}
 \epsscale{0.4}
 \plotone{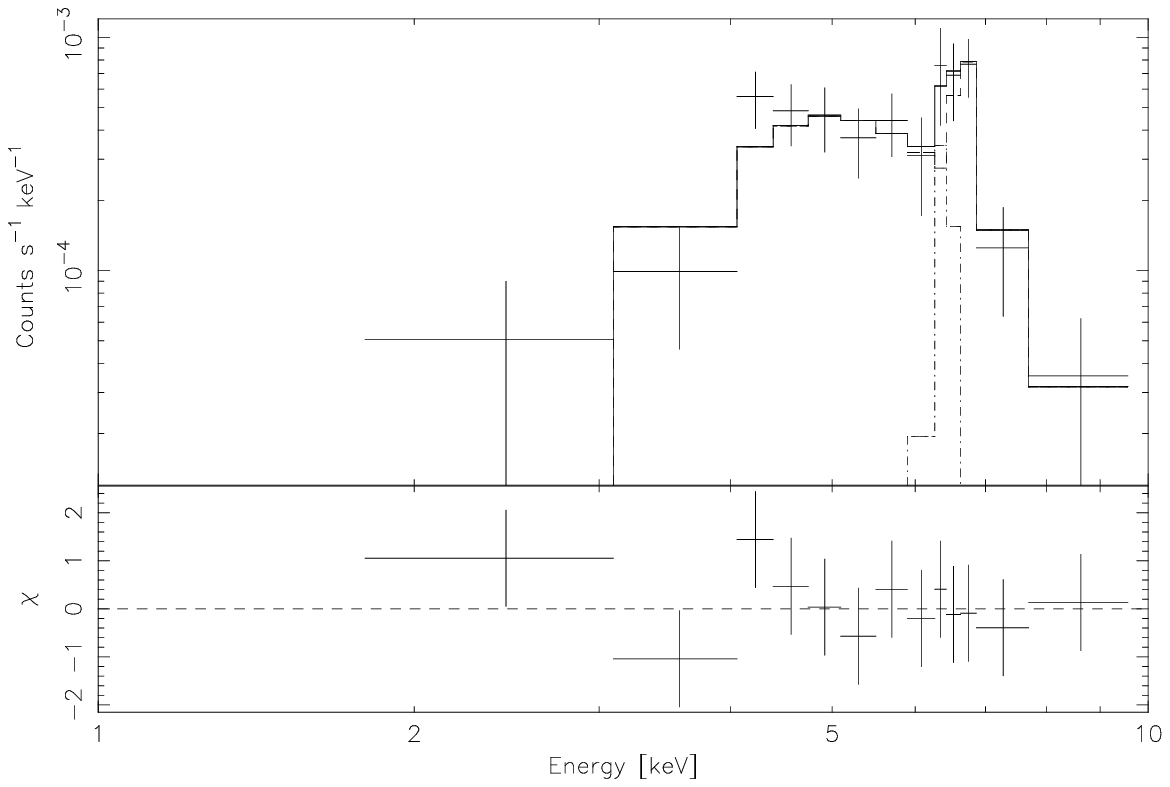}
 \plotone{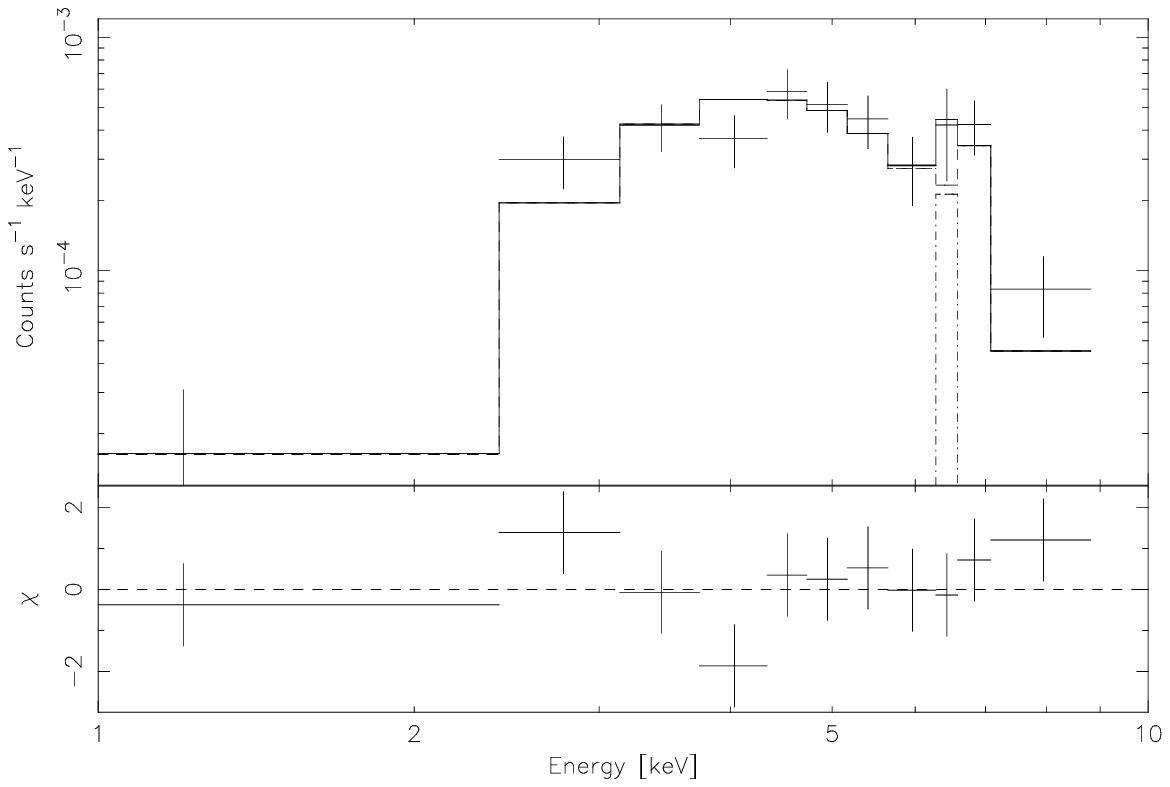}
 \figcaption[f3a.eps,f3b.eps]
{The composite  X-ray spectra of the class~A~(left)  and B~(right) sources
and the best-fit
model of  a thin thermal plasma~(solid histogram) with a 6.4-keV line
~(dotted histogram).
 X-ray photons are extracted either from the box~(for No.~11) or
 circles~(the other sources) with a diameter of $3\times$HPD~(see
 Figure~\ref{fig;b2all}). \label{fig;tashi}}
\end{figure}

Since the statistics of individual  cloud  member is limited (6--80
photons, see Table 2),
we group them into two classes; class~A comprises  No.~10--13, which are  
associated with the HII regions,
and the class~B sources  have  neither radio nor IR counterpart (No.~1,
3, 4, 6--9, and 14--17).  
We then make the composite spectra
for the class~A and B sources (Figure~\ref{fig;tashi}) separately,  
and examine the average properties.   

The composite spectra are nicely fitted with a thin thermal plasma 
of solar-abundances.
However, an excess near at 6.4~keV is found, 
hence we add a  narrow line at 6.4 keV, which represents the K-shell 
transition of neutral or low-ionization irons.
The best-fit parameters and models are shown
in Table~1 and Figure~\ref{fig;tashi}, respectively.
The mean $N_{\rm H}$ of 3.6 and 1.3 $\times 10^{23} \rm ~H~cm^{-2}$ for the
class~A and B sources are consistent with being in the cloud.

\begin{figure}
 \figurenum{4} 
 \epsscale{0.4}
 \plotone{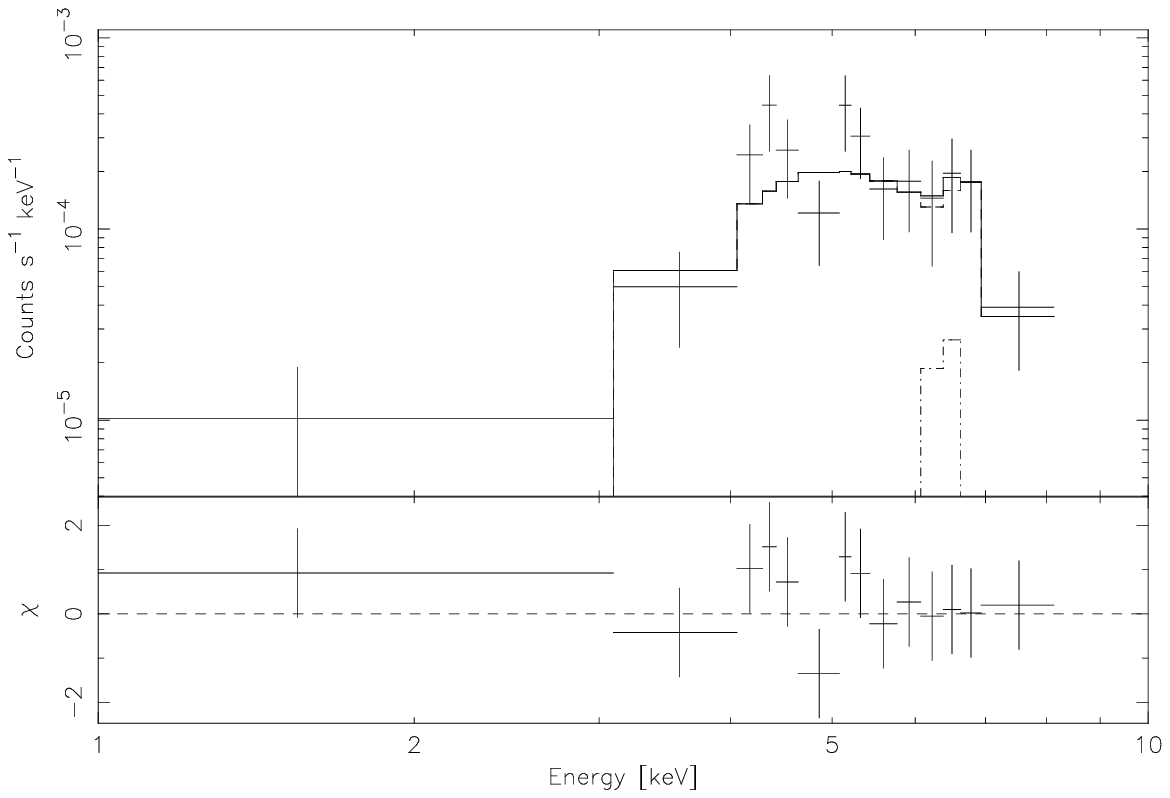}
 \plotone{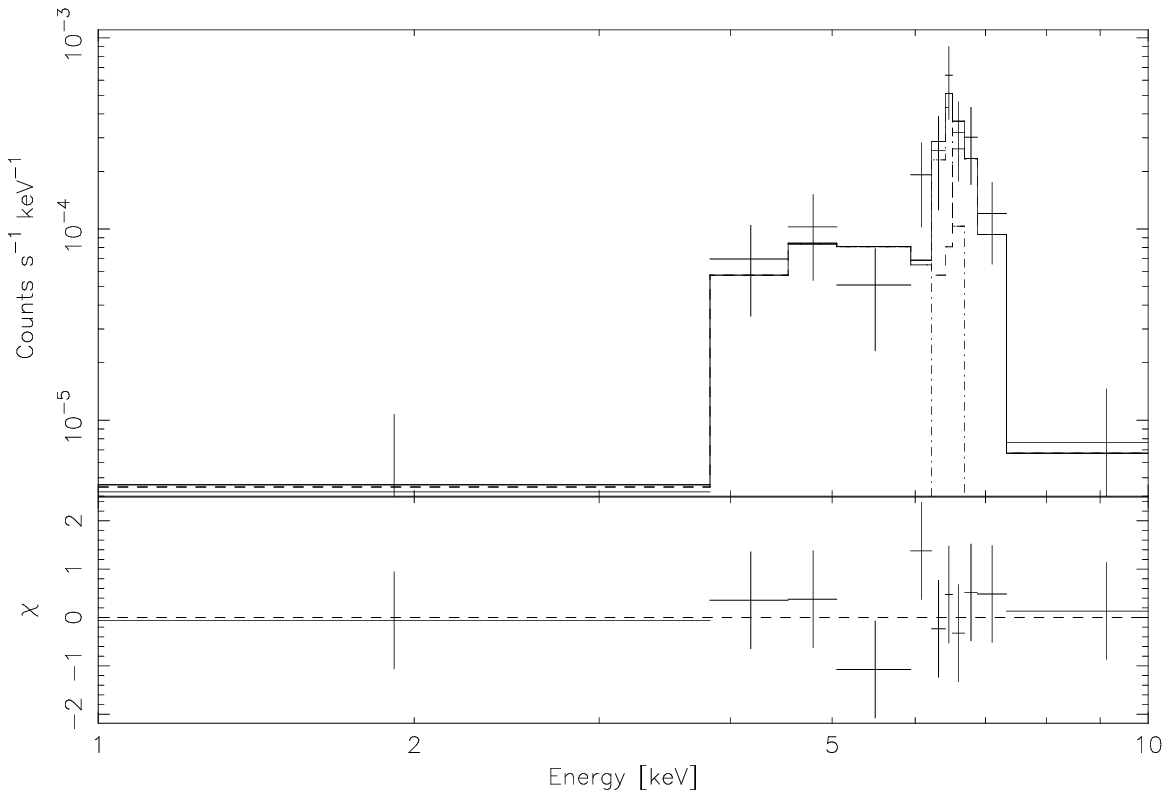}
\figcaption[f4a.eps,f4b.eps]
{The X-ray spectrum and the best-fit model of
a thin thermal plasma~(solid histogram) with a 6.4-keV line~(dotted
 histogram) for sources No.~13~(left) and No.~10~(right).
\label{fig;mainspec}}
\end{figure}

For the spectral fitting of individual X-ray sources, allowable (free) parameters
depend on the continuum and line fluxes.
Since No.~10 shows the strong iron lines (see Figure~4) and the
total flux is reasonably high (51 counts),  we allow 
the temperature ($kT$), absorption column density ($N_{\rm
H}$), abundance ($Z$), and equivalent width
($E.W.$) of the 6.4-keV line to be free.
For comparison, we also use the same free parameters  for  No.~13.
For the other sources, we refer the best-fit parameters
of the composite spectra with free parameters depending on the flux; 
those with flux between 29--50 counts (No.~8 and 11), 
free parameter is $N_{\rm H}$;
and the faintest sources less than  21 counts,
we fix all the parameters
to those of the composite spectra and estimate the luminosity ($L_{\rm x}$)
only.
The best-fit thin thermal plasma model parameters are listed 
in Table~2, while the spectra and the best-fit models 
of the two brightest sources, No.~10 and 13, are shown in
Figure~\ref{fig;mainspec}.

Since the 6.4-keV line feature in addition to the 6.7-keV line may be
sensitive to  the charge transfer inefficiency (CTI), 
we examine the CTI correction 
issue  using  the software described by the method of Townsley~et~al.~(2001).
In order to achieve the best statistics, we use the  diffuse emission from all
the  $3'\times3\arcmin.5$  region  which includes the strong 6.4-keV
line in Sgr~B2 (Murakami et al.~2001).
The standard level~2 process gives the center energy of
$6.40~(6.36-6.41)$~keV, flux of $5.6~(5.1-6.2) \times 10^{-5}~\rm
photons~cm^{-2}~s^{-1}$, and width of $60~(8.8-130)$~eV,
which are essentially the same as those by the method of Townsley~et~al.~(2001)
(see Table~2 in Murakami~et~al.~2001).
Thus we conclude that the Level 2 process correct
the CTI reasonably  well, at least for the present data set.

The 6.4-keV line flux  is also sensitive to the background subtraction, 
because the background itself is a luminous 6.4-keV line source.  We 
therefore re-subtract 
the background  taken from the right half region of Sgr B2, where 
the 6.4-keV line is  about two times larger than the average.  Still
we obtain consistent results. Thus the 6.4- and 6.7-keV iron lines
are not artifacts due to CTI nor background subtraction. 

We examine  the time variability for all sources with the
Kolmogorov-Smirnov test
(Press et~al.~1992) under the constant flux hypothesis  using {\it lcstats} in
the
{\bfseries XRONOS} (Ver 5.16) package\footnote{see
http://xronos.gsfc.nasa.gov}. 
Five (cloud members) and two (non cloud members) sources are 
found to be time variable with more than 90 \% significant level. 
However, no flare-like event with a fast rise and slow decay profile
is found from any of them.

\section{Discussion} 

\subsection{X-rays Associated with the HII complexes} 

No.~10 and 13, the most obvious cloud members, are associated
with the complex of ultra compact HII regions (UC HIIs),  Sgr~B2~(M)
(e.g., Gaume \& Claussen 1990) (see Figure~2). The absorption column 
density of $\sim4\times10^{23} \rm ~H~cm^{-2} $  supports that  
No.~10 and No.~13 are lying in the core of the Sgr~B2 (M).

No.~10 seems to be elongated
along the UC HIIs F3, F4 and G (De pree et~al.\ 1998) and the MIR
source MSX5C~G000.6676-00.0355 (Egan~et~al.~1999), hence would 
comprise several point sources.   
The X-ray peak comes near the position of the brightest UC HII, F3
(De Pree et~al.\ 1998), which is also considered to be an engine of 
the HC$_3$N bipolar flow (Lis et~al.\ 1993).  
From the HII and outflow data, this X-ray peak may correspond to a very
high-mass (e.g., O6, assuming the zero age main sequence (ZAMS) star:
De~Pree et~al.\ 1998), young (well below $10^5$ year: Lis et~al.\ 1993) star.

A notable feature of No.~10 is strong lines  at 6.7~keV
and 6.4~keV.
The best-fit iron abundance of 5 times of solar  is extremely large but, 
within the large error, it is consistent with that of the 
Sgr~B2 cloud of 2--3, 
estimated  from X-ray reflection nebula (Murakami et al.\ 2001)
and that predicted from the abundance  gradient of HII regions~toward the GC  
(e.g., Simpson et~al.\ 1995).
The abundances, on the other hand, is  
significantly larger than that of any other
high-mass SFRs (Kohno~et~al.~2002; Schulz et~al.\ 2001;
Yamauchi et~al.\ 1996). 
It may be noted that He66$\alpha$ line observation by De Pree et~al.\
(1996) predicts anomalously low helium-abundance from the HII region 
F only, in contrast of apparent iron over-abundance with the present study.

No.~13  is  also extended, lying beyond the east filament of the
cometary  HII region  I (Gaume \& Claussen 1990).
The ionized gas filament is likely produced by an interaction of outflow(s)
from the UC HIIs F as is supported by the extremely broad H66$\alpha$ recombination
lines (De~Pree et~al.\ 1996). 
However the outflow velocity
of  $\sim$ 100 km s$^{-1}$  (FWHM of the line width; Gaume
et~al.\ 1995) is insufficient to produce the 5-keV plasma we observed.

The best-fit iron abundance of  No.~13, in contrast to No.~10, is
sub-solar (0.3 of solar), although the  error is large for a concrete  
conclusion.
It is debatable  whether the chemical abundances  or 
the X-ray emission  mechanisms is  different between the two
sources (No 10 and 13)  in Sgr B2 (M)  separated by  only 0.2--0.3 pc.

No.~11 is  extended over the whole region of  Sgr~B2~(N).
The best-fit absorption of 5.7$\times 10^{23} \rm ~H~cm^{-2}$
supports that No.~11 is located in the  core of the UC HII complex.
The radio continuum  and molecular bipolar flow data indicate that 
Sgr~B2~(N) is younger than Sgr~B2~(M) (Gaume et~al.\ 1995; Lis et~al.\ 1993)
and may contain $\sim7$ massive stars ranging from O5 to B0 (Gaume et~al.\
1995).
The He66$\alpha$ line from one of the UC HII (K6) in Sgr~B2~(N) shows 
doubly peaked  profile, indicating an expanding shell of
the dynamical age  of $\sim10^4$ year (De Pree et~al.\ 1995).  
From the expansion velocity of $\sim 22 \rm ~km~s^{-1}$ 
(De Pree et~al.\ 1995), this shell can not be an origin of the hard 
X-rays of No.~11. 

No.~12, at the Sgr~B2~(S), is not resolved, and  is
likely to associated with an isolated star of spectral type O6.5 (Gaume et~al.\ 1995).
The MIR source MSX5C~G000.6564--00.0411 is also found at No.~12 
within $1\sigma$ position error (Egan~et~al.~1999).

\subsection{Origin of the X-rays from  the HII Regions} 

As we discussed in the previous sections, X-ray sources No.~10--13  
are closely associated 
with the UC HII complexes  Sgr~B2~(M), (N) and (S)
with the absorption corrected luminosity (in 2--10~keV)  
of 23, 12 and  $2.2 \times 10^{32}$ erg s$^{-1}$, respectively.
Since  Sgr~B2~(M), (N) and (S)  may contain  $\sim$ a few to 20 UC HIIs 
 each  (De~Pree et~al.\ 1998), No. 10--13  are possibly  clusters 
of  high-mass YSOs  with  similar numbers.  The individual
YSO luminosity is then $\sim10^{32}$ erg s$^{-1}$, which is consistent with the high-mass
YSOs (Kohno et~al.\ 2001; Schulz et~al.\ 2001).

The plasma temperatures of about 5 keV are, however, significantly  higher 
than that of stellar-wind-driven X-rays found in high-mass main sequence
stars (MSSs) (Bergh\"ofer et~al.\ 1997).
We accordingly propose a scenario that high-mass
stars produce  hard ($\ge$5 keV)  X-rays due to the magnetic activity in a
phase of pre main sequence (Babel
\& Montnerle 1997; Hayashi, Shibata, \& Matsumoto 1996).  
In fact, high temperature plasma has been found from high-mass YSOs
in the Orion Nebula and Mon R2 cloud  (e.g., Shultz
et~al.\ 2001; Kohno et~al.\ 2001).
It may continue until ZAMS, then gradually decline to the stellar
wind dominant activity.
Finally, high-mass MSSs predominantly emit soft ($\leq$1~keV  temperature)
X-rays originated from  the strong stellar wind. 

The observed temperature of about 5 keV is, on the other hand, more typical 
to low-mass protostars  (Kamata et al.~1997; Koyama et al.~1996a; 
Tsuboi et~al.~2000;
Imanishi, Koyama, \& Tsuboi~2001) with 
the luminosity of $\sim 10^{28-30} \rm~erg~s^{-1}$ 
(Feigelson et~al.\ 1993; Casanova et~al.\ 1995), 
hence it is also possible that No.~10--13 is a cluster of numerous number
($10^{2}$--$10^{4}$) of low-mass YSOs. The most probable scenario, therefore,
would be  that the X-rays from the UC HII complexes are the  mixture of YSOs
with various mass range (high-medium-low mass stars).

The best-fit  absorption of $\sim 5\times 10^{23}$ H cm$^{-2}$ for 
the Sgr~B2 (M) X-rays is the largest among the known stellar X-ray sources. 
Our observation demonstrates that hard X-rays are very powerful to discover
deeply embedded YSOs  even if they are suffered with a large optical 
extinction of $A$v$\sim$200--300 $mag$.

\subsection {X-ray sources with no HII region  (class~B)} 

In the Sgr B2 cloud, we find 11 class~B sources.
The lowest flux of those sources is estimated to be $3 \times 10^{-15} \rm
~erg~cm^{-2}~s^{-1}$ after removing the absorption of
$1.3 \times 10^{23} \rm ~H~cm^{-2}$, the mean value of the class~B sources.
The expected  number  of extra galactic sources above the flux of 
$3 \times 10^{-15} \rm ~erg~cm^{-2}~s^{-1}$ is  only $\sim 2.5$ in the 
Sgr B2 region of  $3\arcmin \times 3\arcmin.5$ (Mushotzky et~al.\ 2000).
 If we  fit the composite class B spectrum with a power-law
model, the  photon index is found to be  $\Gamma < 1.3$, which is out side
of  the typical value for AGNs (1.5--2.0).
The average $N_{\rm H}$  of $ 1.3 \times10^{23}$ ~H~cm$^{-2}$  also supports that
most of the class ~B sources  are not background AGNs,  but are the cloud members.  
The temperature of $> 4.2 $~keV (see Table~1),
and each X-ray luminosity of a few times  $10^{31-32}$~erg~s$^{-1}$  are
consistent with  those of class~A sources, hence support that  the
class~B  source is either a single or cluster of YSO(s) with various mass ranges. 
The time variable sources is likely  a  single star, either  a  
high-mass or a low-mass in the high end of the luminosity function.

A problem of the YSO
scenario is that no HII region, maser source, nor outflow has been found.
Furthermore,
the star forming regions and high density regions elongate
north to south in Sgr~B2 (e.g., Sato et~al.\ 2000),  while  the class~B
sources are widely distributed over the whole Sgr B2 cloud.
This leads us to an alternative scenario that  isolated
white dwarfs (WDs) are powered by the Bondi-Hoyle accretion (Bondi \& Hoyle
1944) from the dense cloud gas.
In fact, 
the observed thin thermal plasma with the mean  temperature of $>4.2$ keV is
consistent with those of accretion powered  WDs (e.g., Ezuka \& Ishida 1999 and
 references there in). 
The Bond-Hoyle accretion rate is given by;
\begin{equation}
\dot{M} = 2\times10^{15}~(\frac{n_{\rm H_{2}}}{10^4~{\rm cm^{-3}}})~
(\frac{v}{10~{\rm km~s^{-1}}})^{-3}~(\frac{M}{M_{\odot}})^2 \rm ~g~s^{-1},
\end{equation}
where $M$ and $v$ are the mass and traveling velocity of WDs relative to
cloud, and
$n_{\rm H_{2}}$ is the number density of molecular hydrogen  in the cloud. 
The mean $n_{\rm H_{2}}$ can be estimated to be $\sim 10^4$
cm$^{-3}$, from the $N_{\rm
H}$ value of
$10^{24}$ H~cm$^{-2}$ and typical depth of the cloud of about 20 pc
(Lis \& Goldsmith 1989).
The X-ray luminosity is given by
$L_{\rm x} = GM\dot{M}/R$, where $R$ is the radius of the WD and
$G$ is the gravitational constant.
Assuming  the standard mass to radius relation
of
WDs ($R = 5.5\times10^8(M/M_{\odot})^{-1/3}$),  $L_{\rm x}$ is estimated
\begin{equation}
L_{\rm x} = 10^{32}~(\frac{n_{\rm H_{2}}}{10^4~{\rm cm^{-3}}})~
(\frac{v}{10~{\rm km~s^{-1}}})^{-3}~(\frac{M}{0.6~M_{\odot}})^{10/3} \rm
~erg~s^{-1},
\end{equation}
where, $0.6~M_{\odot}$ is the typical mass of WDs (Ramsay 2000 and
references there in).
Accordingly, with a reasonable range of gas density and WD parameters,
the predicted X-ray luminosity is consistent with that of the observations.
The number density of WDs in the solar
neighbor is  (5--6)$\times10^{-3}$ pc$^{-3}$ (Holberg, Oswalt, \& Sion
2001).
 Since the number
density of stars near the  Sgr B2 region (100~pc from the GC)
is more than  $10^2$ times higher
than that the solar neighbor (e.g., Oort 1977), the population of WDs
in our observed area  (assuming a rectangle of 7.2 $\times$ 8.4 $\times$ 20~pc$^3$)
can be estimated to be $>$~600.  Therefore it is conceivable that  the
class~B
sources are a part of isolated WDs, which may lie in the higher end of the
luminosity function.

\subsection{Cloud Cores Irradiated by X-ray Sources} 

The composite spectra of all the cloud members
has a 6.4-keV line with the  $E.W.$ of 200--230 eV. 
The line energy corresponds to the K-shell emission from neutral 
or low-ionized irons, hence the most plausible origin is fluorescence 
from cold irons in the cloud gas.
Assuming  that homogeneous circumstellar gas is  irradiated by a 
central X-ray sources,  the required column density 
to emit observed 6.4 keV line is estimated from the relation of,
\begin{equation}
 E.W. \sim 100~Z_{\rm Fe}~(\frac{N_{\rm H}}{10^{23} {\rm ~H~cm}^{-2}})~\rm
eV,
\end{equation}
where $Z_{\rm Fe}$ is the iron abundance relative to solar (Inoue
1985).
Most of the source photons (except for No.~11) are extracted from a circle
with a $3\times$HPD-diameter, which corresponds $\sim0.1 \rm ~pc$ in radius.
Thus the number density of atomic hydrogen ($n_{\rm H}$) in the
source regions must be 6$\times 10^{5} \rm ~cm^{-3} $,
assuming $Z_{\rm Fe}$ is 1. 
This estimated value exceeds the average $n_{\rm H_{2}}$ in the Sgr~B2 cloud  of
 $\sim 10^{4} \rm ~cm^{-3}$ (see previous subsection).
We further find that the $E.W.$ does not decrease as we reduce the radius
of the photon extraction circles.
Therefore the gas
density should be concentrated very close to the X-ray sources.
In fact, De Pree et~al.~(1998) suggested that the density ($n_{\rm H_{2}}$) in the 
close vicinity ($\sim
10^{-3}$~pc)  of the UC HII region in Sgr~B2~(M) is in between $2 \times 10^{7} \rm cm^{-3}$ 
and $1 \times 10^{8}
\rm cm^{-3}$, if they are in pressure equilibrium.

No.~10 requires exceptionally strong  6.4-keV line with the $E.W.$ of
630~eV.
Using the  $n_{\rm H_{2}}$ value proposed by De Pree
et~al.~(1998) and the core radius of $10^{-3}$~pc, the column
density in the core region of  Sgr~B2~(M) is (1--6)$\times
10^{23} \rm
 ~H~cm^{-2}$.  
Then the expected abundance $Z_{\rm Fe}$ is $\sim$ 1--6, which is
consistent with the best-fit value of a thin thermal fit (see 3.1).

The X-ray sources in  a dense cloud cores such as Sgr~B2~(M),
(N), and (S),  may  photo-ionize the core regions. 
Using the X-ray ionization rate by Lorenzani \& Palla (2001), we estimate 
that a dense core  of  $n_{\rm H} \sim 10^{7-8} \rm~cm^{-3}$ is
largely ionized within  $\sim 0.01-0.03$~pc in radius from  a  
central source of 10-keV  plasma with the luminosity of $10^{32} \rm ~erg~s^{-1}$.  
For Sgr B2 (M), which  may contains  $\sim 20$ X-ray sources in  a $\sim$ 0.15-pc radius,  
the ionized volume is   $\sim 1-20$\% of the core.
Since the ionized gas strongly couples to the magnetic field, it may  have a significant 
effect on the cloud compression for star formation.

The X-ray ionization may also have an impact on the  abundance of molecules. 
The abundance of $\rm HC_{3}N$, for example, drops very rapidly  with 
increasing  ionization.  The $\rm HC_{3}N$ abundance in Sgr~B2~(M) is 20 
times less than that of  Sgr~B2~(N) (De~Vicente et al.~2000). Since Sgr~B2~(N)
is less X-ray active and more largely extended than Sgr~B2~(M), the abundance
difference between these two cloud cores may  be explained with the X-ray ionization effect.

\section{Conclusion} 

\begin{enumerate}

\item  We discovered a dozen X-ray sources in the giant molecular cloud
Sagittarius~B2 (Sgr~B2).
Judging from the large absorptions ($N_{\rm H}$) of
(1--10)$\times10^{23}$~H~cm$^{-2}$ and log$N$-log$S$ relation for the 
background sources, most of these are  the cloud members.

\item  The brightest two sources are found near at the HII complex  Sgr~B2~(M),
which are likely the
cluster of X-ray emitting YSOs.  The X-ray spectra are largely
different between
the two sources. One has extremely strong iron lines, while the other has
only weak lines.

\item  X-ray excess from the HII complexes Sgr~B2~(N)  and (S) are found,
the former is extended, while the latter is point-like.  Both have the X-ray spectra 
and luminosities consistent with YSOs.  No significant X-rays are found from any 
other HII regions.

\item  The other  X-ray sources also have the X-ray spectra and luminosities
consistent with a single or a cluster of YSO(s).  
However, no counterpart of any other
wavelength,  neither optical/IR star, HII region, maser source nor
molecular core has been reported. 
Thus alternative possibility is isolated white dwarfs powered by
the Bondi-Hoyle  accretion.

\item  Strong 6.4-keV lines are found from the X-ray sources. 
The origin of this line would be the fluorescence of the circumstellar 
gas irradiated by the central stars.  
The gas density is not uniform but is centrally  concentrated.

\end{enumerate}

\vspace*{1em}

\acknowledgements
We thank to our referee, E. Feigelson for his useful comments and suggestions.
H. M. is financially supported by the Japan Society for the Promotion of
Science.

\onecolumn
\newpage

{
 \begin{deluxetable}{lcccccc}
 \tabletypesize{\scriptsize}
 \tablenum{1}
 \tablecaption{BEST-FIT RESULTS OF THE COMBINED SPECTRA
  }
 \label{tab:tashi}
 \tablehead{
 \colhead {class} &\colhead{$N_{\rm H}$}
 &\colhead{$kT$} &\colhead{$Z$}
 &\colhead{$E.W.$}
 &\colhead{$L_{\rm X}$}&\colhead{Reduced~$\chi^2~(d.o.f)$}
 \nl
 \colhead{} &\colhead{ (1)}&\colhead{ (2)}
 &\colhead{ (3)} &\colhead{ (4)} &\colhead{ (5)}
 &\colhead{}
 }
\startdata
 A &{3.6~(2.8--4.5)}  &{6.5~(2.9--13)} & {1.0~(fixed)} &{200~($<420$)} &
 {33~(26--38)}&{0.6~(9)}
\nl
B &{1.3~(0.93--1.8)}  &{9.5~($>$4.2)} & {1.0~(fixed)}  &{230~($<480$)} &
{15~(13--17)} & {1.0~(8)}
\nl

\enddata
\tablecomments{Parentheses indicate the 90\% confidence limit.
(1):absorption column
 density of hydrogen ($10^{23} \rm ~H~cm^{-2}$). (2):plasma
 temperature (keV). (3):abundance (solar ratio). (4) equivalent width (eV)
of the
6.4-keV line (Fe~$K_{\alpha}$ of neutral iron).
(5):absorption corrected luminosity in 2--10~keV ($10^{32} \rm
~erg~s^{-1}$).
}

\end{deluxetable}
}

\begin{center}
{
\begin{deluxetable}{lllcrlllllll}
\tabletypesize{\scriptsize}
\tablewidth{0pt}
\rotate
\tablenum{2}
\tablecaption{{\it Chandra} X-RAY SOURCES IN THE SGR~B2 CLOUD
}
\label{tab:naka}
\tablehead{
\colhead{No.~} & \colhead{R.A.} & \colhead{Dec.} & \colhead{Cts} &
\colhead{Var.}&
 \colhead{$N_{\rm H}$} & \colhead{$kT$} & \colhead{$L$x} & \colhead{$Z$} &
 \colhead{$E.W.$} & \colhead{counterpart} &\colhead{size}\nl
 \colhead{} & \colhead{(J2000)} & \colhead{(J2000)} & \colhead{(1)} &
 \colhead{(2)} & \colhead{(3)} & \colhead{(4)} & \colhead{(5)} &
 \colhead{(6)} & \colhead{(7)} & \colhead{(8)} & \colhead{(9)}
}
\startdata

1  & 17:47:13.6  & $-$28:23:37.8 & 19 & \nodata & --$^{\dagger}$
   & -- & 0.83~(0.51--1.2) & --  & -- & \nodata & $ 2.\arcsec 38$ \nl

2  & 17:47:14.1 & $-$28:23:31.1  & 13 & $>99.9$     & 0.56$^{\star}$~(0.34--0.90) 
   & \nodata &  (5.7$^{\star \star}$~(4.1--7.3)) & \nodata  & \nodata& (foreground) & $ 2.\arcsec 59$   \nl

3  & 17:47:15.1 & $-$28:24:41.6 & 7.2 & \nodata & -- 
   & -- & 0.40~(0.14--0.65) & --  & -- & \nodata & $ 1.\arcsec 65$ \nl

4  & 17:47:15.5  & $-$28:21:40.8 & 15 & 99.2    & --  
   & -- & 1.2~(0.68--1.7)   & --  & -- & \nodata & $ 5.\arcsec 44 $ \nl

5  & 17:47:15.8  & $-$28:24:17.1 & 18 & $>99.9$    & 0.082$^{\star}$~(0.059--0.11) 
   & \nodata & (3.8$^{\star \star}$~(2.9-4.7)) & \nodata  & \nodata & (foreground) & $2.\arcsec 01$ \nl

6  & 17:47:16.4  & $-$28:21:32.2 & 18 & \nodata & -- & -- & 0.83~(0.50--1.2)
& --  & -- & 
\nodata & $ 6.\arcsec 00$
   \nl

7  & 17:47:17.2  & $-$28:24:09.2 & 12 & 99.5   & -- & -- & 0.50~(0.25--0.75) & -- &
-- & \nodata  & $2.\arcsec 40$ \nl

8  & 17:47:18.2  & $-$28:23:48.7 & 41 & \nodata & 0.78~(0.49--1.4) & -- & 2.2
~(1.5--2.7)
& --  & -- & \nodata  & $ 3.\arcsec 00$
   \nl

9  & 17:47:20.1  & $-$28:24:16.9 & 14 & 96.6 & -- & --  & 0.65~(0.35--0.94) & --
& --  & \nodata & $ 2.\arcsec 96$
   \nl

10 & 17:47:20.2  & $-$28:23:05.3 & 51 & \nodata & 4.0~(1.9--8.8) & 10~(4.1--30)
& 8.7
~(4.3--17)
   & 5.0~($>$0.94) & 630~(180--1100) & Main & $ 4.\arcsec 24$
   \nl

11 & 17:47:20.3 & $-$28:22:15.3 & 29 & \nodata & 5.7~($>$1.8) & --  & 12
~(5.3--18)
   & --  & -- & North & $ 25.\arcsec 0 \times 21.\arcsec 0 ^{\ast} $
   \nl

12 & 17:47:20.4 & $-$28:23:45.4 & 12 & \nodata & -- &
--  &
 2.2~(1.1--3.3) & --  & -- & South & $ 4.\arcsec 00$ \nl

13 & 17:47:20.9  & $-$28:23:06.2 & 83 & \nodata & 3.8~(2.3--7.2)  & 5.1
~($>$1.1) & 14
~(11--18)
   & 0.31$^{\ddagger}$ & 62~($<$370) & Main & $ 5.\arcsec 20$
   \nl

14 & 17:47:22.3  & $-$28:23:26.4 & 15 & 95.6 & -- & -- & 1.4~(0.79--2.0) & --
 & -- &  \nodata & $ 4.\arcsec 35$
   \nl

15 &17:47:23.1  & $-$28:22:32.4  & 20 & 91.0 & --  & --  &
 1.3~(0.81--1.8) & -- & -- & \nodata & $ 5.\arcsec 72$\nl

16 & 17:47:23.2  & $-$28:23:25.3 & 6.0 & \nodata 
 & -- & -- & 0.26~(0.080--0.44) & -- &
-- & \nodata  & $4.\arcsec 59$ \nl

17 & 17:47:23.8  & $-$28:22:31.7 & 6.0 & \nodata & -- & -- &
 0.66~(0.20--1.1) & -- & --  & \nodata & $ 6.\arcsec 00$\nl

\enddata \tablecomments{
 Parentheses indicate the 90\% confidence limit.
 Col.~(1):photon counts~(background subtracted).
 Col.~(2):variability of more than 90 \% confidence level is given 
with the Kolmogorov-Smirnov test.
 Col.~(3):absorption hydrogen column density~($10^{23} \rm~H~cm^{-2}$).
 Col.~(4):plasma temperature~(keV).
 Col.~(5):absorption corrected luminosity in 2--10~keV~($10^{32} \rm
~erg~s^{-1}$).
 Col.~(6):abundance relative to solar value.
 Col.~(7):equivalent width~(eV) of the 6.4-keV line~(Fe~$K_{\alpha}$ of
 neutral iron). 
 Col.~(8):counterpart of HII complexes (No.~10--13).
 Col.~(9):the radius of photon extracted circle.\nl
 $^{\dagger}$: Hyphen~(--) means the fixed value to the best-fit parameters
of the composite spectra~(see Table 1).
 $^{\ddagger}$:Errors are not constrained.
 $^{\star}$: The fitting model is a 1-keV thin thermal plasma of solar abundance.
 $^{\star \star}$: absorption corrected X-ray flux in 2--10~keV~($10^{-15} \rm~erg~cm^{-2}~s^{-1}$).
 $^{\ast}$: The source region is rectangle.
}
\end{deluxetable}
}
\end{center}

\end{document}